# A Nanoscale Parametric Feedback Oscillator

*L. Guillermo Villanueva, Rassul B. Karabalin, Matthew H. Matheny, Eyal Kenig, Michael C. Cross, Michael L. Roukes**

Kavli Nanoscience Institute, California Institute of Technology, Pasadena, CA, 91125

Email: roukes@caltech.edu

TITLE RUNNING HEAD: A Nanoscale Parametric Feedback Oscillator

ABSTRACT

We describe and demonstrate a new oscillator topology – the parametric feedback oscillator (PFO). The PFO paradigm is applicable to a wide variety of nanoscale devices, and opens the possibility of new classes of oscillators employing innovative frequency-determining elements, like such as nanoelectromechanical systems (NEMS), facilitating integration with circuitry, and reduction in cost and system size reduction. We show that the PFO topology can also improve nanoscale oscillator performance by circumventing detrimental effects that are otherwise imposed by the strong device nonlinearity in this size regime.

MANUSCRIPT TEXT

Frequency stability is essential for self-sustained oscillators that are at the heart of many current technologies spanning communication, computation, and geolocation[1-2]. An oscillator's essential elements are a *frequency-determining element*, typically a mechanical or electrical resonator with a high quality factor response (high *Q*); and a *feedback loop*, usually composed of a linear amplification stage, a signal limiter, and a phase-delay element. Here, by the term *resonator*, we refer to a passive device that requires AC power in order to be driven into motion. In an *oscillator*, the feedback signal provides sufficient drive to overcome the resonator's damping and, thereby, to sustain continuous vibrations.

The most common type of frequency-determining element for oscillators is a macroscopic quartz crystal[3]. Quartz-crystal-based oscillators have been the prevalent standard for almost a century[4], and have remained such despite the semiconductor microelectronics revolution. Recent advances in micro- and nano- fabrication permit miniaturization of *semiconductor*-based mechanical resonators, potentially facilitating their on-chip integration with electronic components. Initial steps towards scaling such integrated mechanical resonators downward to the realm of microelectromechanical systems (MEMS) shows promise[5-6]. Yet full compatibility with very-large-scale integration (VLSI) would require that a resonant element their dimensions should ultimately be reduced even further, to the domain of nanoelectromechanical systems (NEMS[7]), so as to become directly compatible in size with individual transistors.

A complementary motivation for pursuing low-noise nanoscale oscillators emerges from the realm of sensing. NEMS resonators are increasingly being employed for sensing applications ranging from detection of mass[8-9], gas[10-11], biomolecules[12-13] and force[14-15] – and provide unprecedented resolution. The most responsive sensing modalities typically employ frequency-shift sensing, a configuration where the stimulus to be measured induces a change in the resonant frequency of the NEMS. A stable NEMS oscillator element, co-integrated with a few active transistors, thus enables an especially compact sensing "pixel".

Whether for frequency control or sensing, implementing nanomechanical oscillators has proven to be extremely challenging, mainly due to the small magnitude of the motional signal generated by the NEMS in comparison to the parasitic cross-talk from the drive[16]. As the dimensions of a NEMS element shrink, so do both the electrical signal produced by its mechanical motion and its onset of nonlinearity[17]. This makes it exceptionally difficult to harness and control the motional signal produced by the NEMS in the in the face of the unavoidable stray reactances that, generally, cause overwhelmingly large background signals. As has been discussed previously[16], in certain cases these deleterious phenomena can be circumvented by implementing carefully constructed bridging and filtering circuits, but these solutions are not universally applicable nor are they easily integrated on-chip.

In this Letter, we present an alternative to the canonical oscillator topology, namely, a new architecture based on non-resonant *parametric* feedback that can be applied to a wide variety of nanoscale resonators. We use a feedback loop possessing a quadratic transfer function to apply parametric excitation at twice the resonant frequency (as opposed to “direct drive” at its resonant frequency). This parametric drive sets the resonator in motion by dynamically modulating one of its physical parameters[18-19]. A complete mathematical analysis of parametric feedback oscillators is provided in the Supporting Information.

As we show, there are many advantages of this generalized parametric feedback technique, both from fundamental and technological points of view. Among them are: (a) it becomes possible to circumvent the need to satisfy the “Barkhausen criteria” that govern conventional oscillators[20]. This makes it possible to harness a wide variety of nanoscale resonators that would otherwise be impossible to employ for oscillator circuits. (b) An unprecedented level of control of the resonator’s nonlinear characteristics is afforded, enabling access to higher amplitudes of operation. (c) A wide range of frequency tunability becomes achievable. (d) Substantial improvement in frequency stability of the oscillator, compared with that of conventional direct-drive implementations, becomes possible.

Here we demonstrate the PFO concept using piezoelectric NEMS. We pattern doubly-clamped beam NEMS resonators from a four-layer stack of aluminum nitride (AlN)-molybdenum (Mo)-AlN-Mo, having a total thickness of 210 nm, a width of 470 nm and a length of 9 μm (Fig. 1a). Our fabrication process is described in detail in the Section I of the Supporting Information. We use this materials combination because it enables the fabrication of NEMS resonators with easily accessible and analytically predictable nonlinear behavior[17] that can be easily excited directly[21] and parametrically[22] by means of the piezoelectric effect. Such piezoelectric NEMS are promising candidates for future co-integration with chip-based electronic circuitry given their small size and compatibility with CMOS processes.

We detect the out-of-plane resonator motion using the time-varying, strain-induced resistance changes in a piezometallic (Mo) loop patterned at one end of the beam (Fig. 1b). Actuation is obtained by applying an AC voltage to an electrode that covers most of the beam's length; this induces longitudinal strain by means of the inverse piezoelectric effect. This time-varying strain can be used to actuate the beam either directly[21,23] or parametrically[22,24]. By driving the beam directly, we determine the natural frequency ($f_0$ = 14.305 MHz) and quality factor ($Q$ = 1220) of the specific device used in these studies by fitting its driven resonant response to a Lorentzian peak (Fig. 1c-bottom). By separately measuring the thermomechanical noise without any drive we calibrate the piezoresistive response to absolute displacement; from this we deduce the transduction responsivity of 8.7 nm/mV for a constant bias voltage of 200 mV across the piezoresistor (Fig. 1c-top). Our doubly-clamped beam devices exhibit stiffening behavior, characteristic of a Duffing nonlinearity, at large drive levels; we use the deduced transduction responsivity to ascertain that the critical amplitude characterizing the onset of nonlinearity is 9.6 nm. This agrees with the predictions of analytical calculations[17] (Fig. 1d-top). The resonance frequency can be tuned by application of a voltage to the actuation electrode (Fig. 1d-inset). We find a tuning sensitivity of 35 kHz/V for this device. This significant tunability readily enables parametric excitation; we subsequently characterize this by sweeping the drive frequency and monitoring the amplitude of vibration at half the applied drive frequency (Fig. 1d-bottom). Especially noteworthy is the plot at the bottom of Fig. 1d that shows both amplitude and frequency detuning grow faster with drive when the device is actuated parametrically at $2f$ as compared with results when directly driven at $f$ (Fig. 1d-top). For example, at 120 mV drive the amplitude and detuning are higher for direct drive at $f$, but when the drive levels exceed 130 mV the situation is reversed. Accordingly, motion amplitudes for the same driving voltage can be much higher in the parametric case, and this is of special significance when building a low noise oscillator.

The schematic of our implementation of parametric feedback oscillator topology is shown in Fig. 2a. Out-of-plane mechanical motion is transduced by providing a constant DC bias voltage across the piezoresistor. The motional signal is then amplified and filtered to suppress high frequency noise and higher harmonics ($>f$). Subsequently, the signal is delayed by a voltage controlled phase shifter ($\phi$), then passed through a nonlinear element optimized to generate a $2f$ signal with amplitude proportional to the square of the resonator motion (see Supporting Information Section II and VI). This frequency doubled waveform is subsequently passed through a highly selective bandpass filter to ensure strong suppression of undesired harmonic content ($f$, $3f$, $4f$, etc.). This ensures that the drive signal fed-back to the resonator is purely sinusoidal at $2f$, which prevents the induction of undesired motional response. Pure modulation at $2f$ induces a resonator response at $f$ – that is, parametric oscillation – provided the $2f$ drive level exceeds the parametric threshold. This threshold can be surpassed in any resonator that has sufficient susceptibility to parametric tuning to permit its resonance frequency to be shifted by more than twice its linewidth[22]. We have evaluated the feasibility of achieving the parametric threshold for a variety of state-of-the-art resonators in Section III of the Supporting Information. Elements scaled down to the nanoscale in all dimensions attain high frequencies with low force constants; this proves ideal for attaining a low parametric threshold.

Unlike behavior in traditional oscillators, the zero-amplitude state is a stable solution for our feedback system. This makes it necessary to initiate resonator motion by an external “start-up” source. After oscillations commence, this start-up drive can be removed and stable oscillations at $f$ will persist, sustained only by parametric feedback at $2f$. In steady state, the parametrically-driven resonator acts as a frequency divider in the circuit. Given that the frequency of the feedback ($2f$) and output signal ($f$) are well separated for the parametric feedback oscillator, their crosstalk is minimized. This eliminates a traditional obstacle for oscillators based on small mechanical devices; for small electromechanical resonators, the output electrical signal is usually strongly dominated by the feed-through of the actuation voltage. In such conditions it is very challenging to attain an oscillation that uses the mechanical resonator as the frequency determining element.

The equation of motion of our doubly-clamped beam PFO system can be written as:

$$\ddot{x} + \frac{2\pi f_0}{Q}\dot{x} + (2\pi)^2 f_0^2[1+\zeta(t,\phi)]\,x + ax^3 + \eta x^2\dot{x} = G(t)\,. \tag{1}$$

Here $x$ represents the displacement of the resonator; $\alpha$ is the nonlinear spring constant (also called the Duffing parameter); $\eta$ is the coefficient of nonlinear damping; $G(t)$ is an external drive signal ($G$ = 0 when the system is in self-sustained parametrically fed-back oscillation); and $\zeta(t,\phi)$ is the feedback function, which depends on the resonator displacement and the externally controlled phase delay (as shall be described below and in Section II of the Supporting Information).

Detailed analysis of equation (1) shows that by varying the two parameters characterizing the parametric feedback, its phase delay $\phi$ and gain, control of both the resonator's effective nonlinear stiffness (the Duffing coefficient, proportional to $x^3$) and its nonlinear damping (proportional to $x^2\dot{x}$) becomes possible (see Section II of Supporting Information). We demonstrate this experimentally by measuring the driven resonant response with parametric feedback *below* the oscillation threshold, for different values of $\phi$. Sweeps of the driven amplitude, as shown in Fig. 2b-d, display induced changes in the resonator's nonlinear coefficient. This evolves from negative (shown in Fig. 2b) to positive values (shown in Fig. 2d). At an intermediate feedback phase (shown in Fig. 2c), the effective nonlinear Duffing constant vanishes and effective nonlinear damping becomes apparent; this is reflected in the increased peak widths at higher drive levels. This control of the nonlinear properties of the system opens possibilities for combined operation, using both parametric ($2f_0$) and direct-drive feedback ($f_0$), to increase the system's dynamic range and improve its frequency stability.

Rotation of the external parametric feedback phase leads to a direct reduction of the non-linear damping. When this reduction is sufficient, the aforementioned parametric oscillation criterion is satisfied (see Section IV in Supporting Information) and oscillations ensue. We characterize the resulting parametric oscillations by capturing their power spectrum, and compare this to the open-loop resonator frequency response (Fig. 3a). The implementation presented here is one of the few examples of a NEMS oscillator reported to date[16,25-26] and, we believe, represents the first realization of a parametric feedback oscillator in any system. For our prototype PFO, we deduce an effective quality factor of 99,000 from its power spectrum; this is more than eighty times larger than the $Q$ of the NEMS resonator itself when operating in its linear regime.

We now analyze the oscillator behavior as a function of the phase shift $\phi$. In Fig. 3b we plot the spectral response of the oscillator for three different values of $\phi$, each incremented by 10 degrees, which results in a frequency shift increment of about 140 kHz (≈14 kHz/deg). Fig. 3b also shows, for comparison, the open-loop resonator response at 20 mV drive. Our theoretical analysis (Supporting Information, Section II) predicts that oscillation frequency and amplitude should both display a strong dependence on $\phi$. We verify this prediction by experimentally monitoring the oscillation frequency while quasistatically changing $\phi$ (using a voltage-controlled phase shifter). Fig. 3c shows the large tuning range obtained, which is almost 18% (from 14.35 to 16.9 MHz). This wide tuning range should prove useful for applications requiring voltage controlled oscillators[27], and for the potential synchronization of coupled oscillators[28]. The extended tuning range is a direct consequence of using parametric feedback: as was shown in Fig. 1, frequency pulling induced by the parametric drive ($2f$) is much more efficient than that obtained from a direct drive (at $f$). The phase shift range accessible in our experiments is about 1400º, and enables our observation of parametric oscillations on three adjacent branches of the phase response, each separated by 360º (Fig. 3c). Excellent quantitative agreement between theory and experiment is evident.

Detailed inspection of our experimental results reveals a flattening of the frequency versus phase data in one specific region, which deviates from our initial theoretical model. This flattening occurs near 16.2 MHz for our device, and is observed in all three of the branches displayed (Fig. 3c and inset). These features arise from the coupling of the fundamental out-of-plane vibrational mode and its first in-plane mode (~32.4 MHz). By modeling an interaction between these two modes[29] we obtain refined predictions that qualitatively match the experimental findings (Supporting Information, Section V).

To assess the performance of our parametric feedback oscillator as a frequency source, we measure its frequency stability. We measure the oscillator's *phase noise*, which represents the sideband power spectral density at a given offset frequency, normalized by the oscillator's signal power[16]. To provide a baseline for comparison, we separately construct a conventional feedback loop with direct drive at $f$, using the same resonant element and active components. Frequency stability comparisons between this direct-drive oscillator and our prototype PFO, for operation at identical energies, are shown in Fig. 3d. We observe that the frequency stability of our PFO is significantly improved compared to that of the traditional oscillator topology. This provides direct evidence of suppression that the PFO topology can suppress the effects of phase noise in the feedback electronics.

In the inset of Fig. 3d, which shows the PFO phase noise at 1 kHz offset, the relative improvement of the PFO's frequency stability is seen to remain relatively constant over the full range of $\phi$. However, very striking enhancement is observed in the region where mode-coupling occurs; in this regime the oscillator's phase noise is reduced by an additional 15 dB. This enhanced noise suppression is consistent with the fact that the frequency instability of our system is dominated by phase fluctuations in the fed back signal. Since phase noise is proportional to the slope of the phase tuning data, there is less noise associated with such "flattened" regions (see Section II in Supporting Information). We anticipate that substantial improvement in frequency stability, ultimately down to the fundamental thermal noise limit, should become possible with optimal engineering of the frequency-phase dependence by such means. In fact, it should be possible to suppress essentially any noise mechanism that originates within the feedback loop itself. Thus, the PFO topology offers a means for resolving the long-standing challenge of attaining ultimate thermodynamic limits of performance in oscillators.

In this work, we describe and demonstrate a novel oscillator circuit topology that employs a nonlinear, parametrically-actuated NEMS doubly-clamped beam, with feedback characterized by a nonlinear, square-law dependence on resonator signal. The advantages of this architecture, which include elimination of cross-talk, control of non-linear properties, large frequency tunability, and significant phase noise reduction, are evident from the experimental results we demonstrate. Since the requirements for realizing a PFO rely solely on the presence of sufficient frequency tunability, the PFO architecture offers wide applicability and outstanding frequency scalability across a variety of possible implementations. This opens new avenues for realizing miniaturized micro- and nano-scale mechanical oscillators based on resonator technologies ranging from MEMS electrostatic disc to graphene NEMS, and it should facilitate very large scale integration of such oscillators with state-of-the-art electronic circuitry.

ACKNOWLEDGMENTS

This work was supported by the Defense Advanced Research Projects Agency Microsystems Technology Office, Dynamic Enabled Frequency Sources Program (DEFYS) through Department of Interior (FA8650-10-1-7029). We thank R. Lifshitz and X.L. Feng for discussions, and P. Ivaldi, E. Defaÿ and S. Hentz for providing us with the AlN/SOI material. L.G.V. acknowledges financial support from Prof. A. Boisen and the European Commission (PIOF-GA-2008-220682).

SUPPORTING INFORMATION PARAGRAPH

**Supporting Information Available**: Piezoelectric NEMS fabrication process flow, PFO applicability criterion, theoretical considerations (derivation of PFO amplitude equation, PFO oscillation condition, modeling of inter-mode coupling feature), thermomechanical limit for phase noise. This material is available free of charge via the Internet at http://pubs.acs.org.

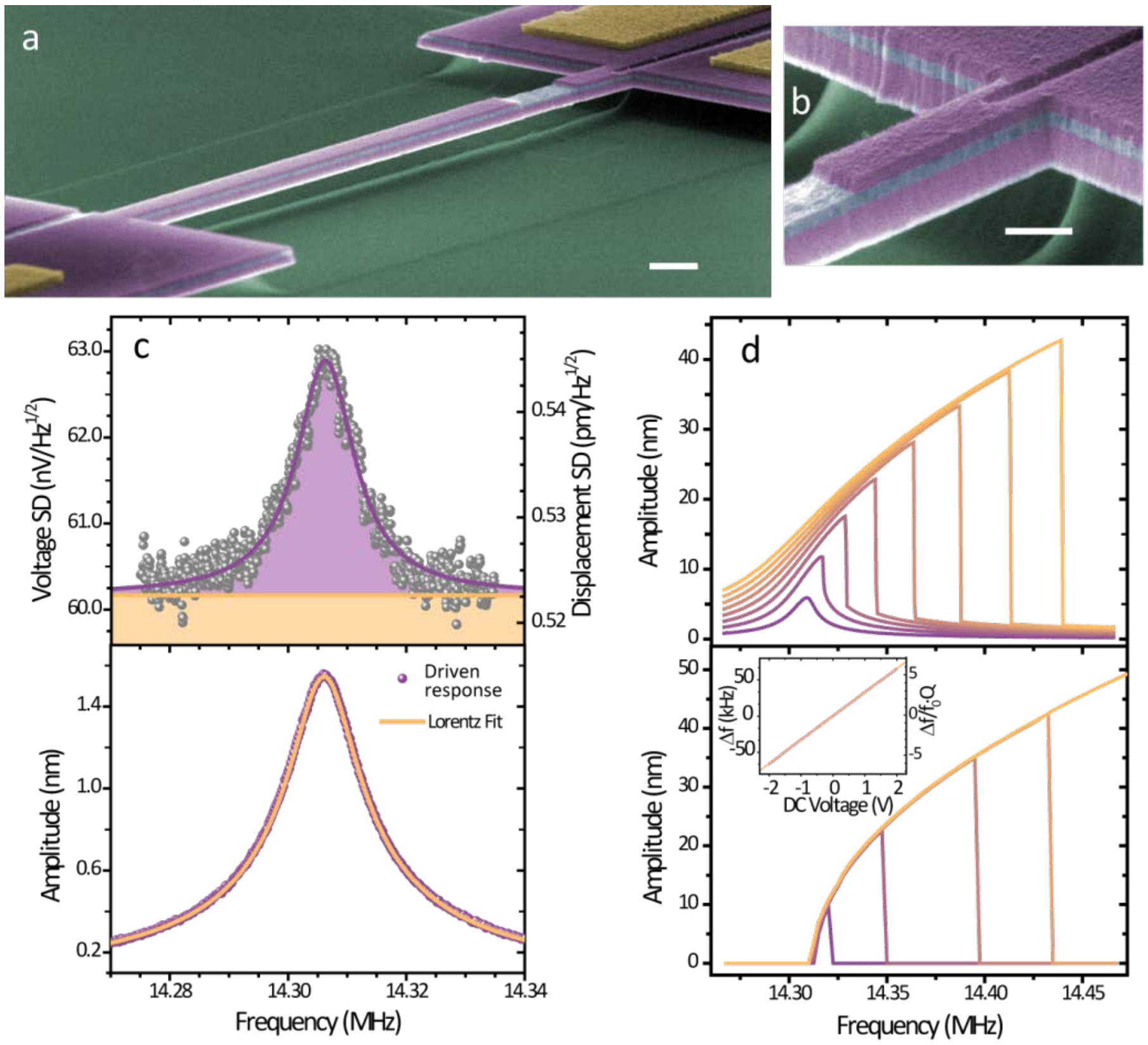

**Figure 1 | Nanomechanical resonator characteristics. a,** Colored SEM micrograph of the suspended mechanical device used to demonstrate the generalized feedback oscillator. The metal electrode covering most of the beam's length is used for actuation, whereas the loop on the opposite side is used for detection. **b,** Detail of the piezometallic loop used to transduce the motion of the resonator. Scale bars: 500 nm. **c,** Top: Voltage spectral density showing the background (system) noise and the thermomechanical peak of the resonator. Detection efficiency (responsivity) of the system is estimated to be 8.7 nm/mV and sensitivity is 0.52 pm/Hz$^{1/2}$. Bottom: Linear resonant response of the resonator in the vicinity of its characteristic resonant frequency. A Lorentzian fit reveals $Q$=1220. **d,** Top: Direct drive of the resonator. Curves show the amplitude response of the resonator around its natural frequency for different driving forces (from 20 mV to 160 mV in steps of 20 mV). A characteristic stiffening effect can be seen and fitted to a Duffing model, to obtain a critical amplitude of about 9.6 nm and a nonlinear dissipation coefficient of 0.015[20]. Bottom: Parametric excitation of the resonator. Curves show the amplitude response of the resonator as a function of half the driving frequency for different driving amplitudes (from 120 mV to 133 mV in 3 mV steps) showing the parametric excitation of the resonator. (Inset) Tunability of the characteristic frequency of the resonator versus DC voltage applied to the actuation electrode (35 kHz/V).

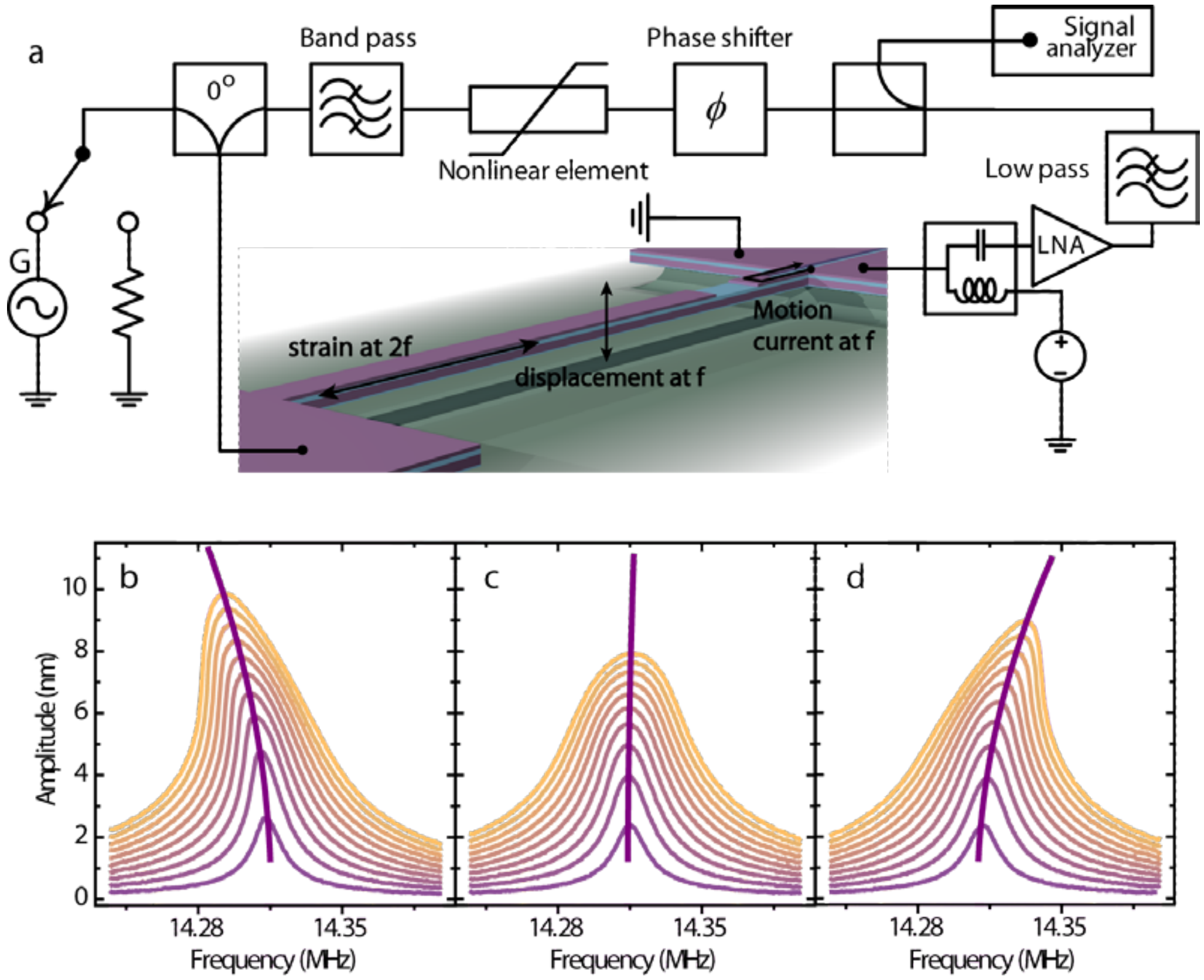

**Figure 2 | Feedback system and results below oscillation threshold. a,** Schematic diagram of our generalized-feedback system. The signal from the resonator is amplified and filtered at high frequencies to eliminate higher harmonics and noise. After an externally controlled ($\phi$) phase delay is applied, the signal is passed through a nonlinear element followed by a bandpass filter to ultimately generate a signal at $2f$. This signal is applied to the actuation port of the beam through a power combiner that allows simultaneous feedback and a direct drive with an external source ($G$), which is necessary to initiate the oscillations. Once the self-sustaining state has been reached, the source can be disconnected and the motion persists. **b-d,** Driven resonant response of the system with feedback *below* oscillation threshold. Amplitude of motion is plotted as a function of drive frequency for 10 different drive values (from 10 to 100 mV in steps of 10 mV). **b-d,** are for different values of $\phi$ (in increments of 45°). The resonator's effective nonlinear coefficient tunes from negative **b,** to positive **d**. In **c,** we show that the nonlinear Duffing coefficient vanishes and the effective nonlinear damping is more apparent.

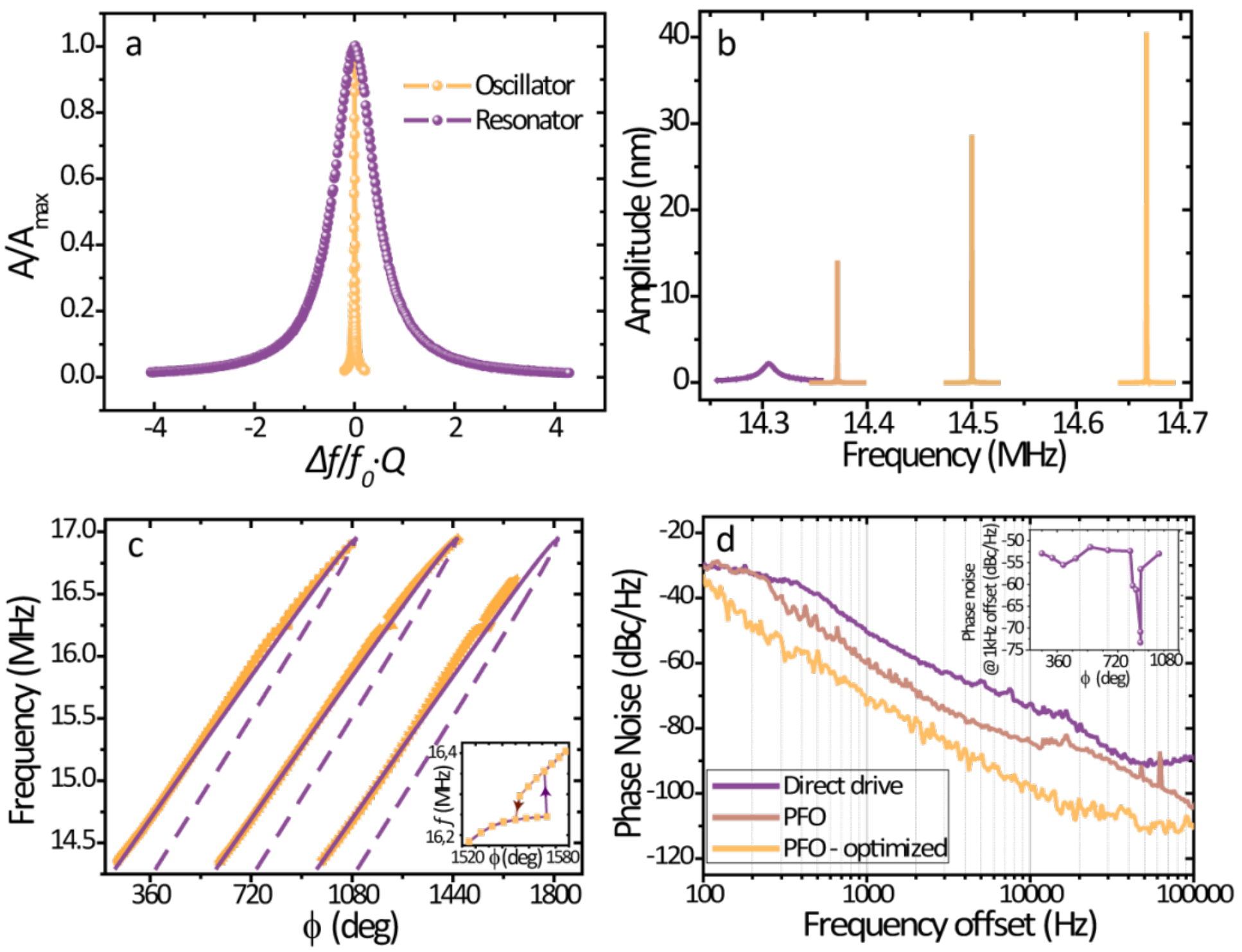

**Figure 3 | Parametric Feedback Oscillator. a,** Normalized comparison between the spectral power of a PFO (orange) and the linear resonant response of the open-loop system (purple). The compression ratio is approximately 82. **b,** Comparison between the open loop response (purple) and the PFO power spectrum (orange). Three PFO traces are shown, each for a different value of $\phi$ (separated by 10$^o$). The tunability in this frequency range is around 14 kHz/$^o$. **c,** Dependence of PFO frequency on $\phi$. Three sets of data (orange) are experimental measurements corresponding to three different solution branches, separated by 360$^o$. Theoretical predictions (purple) showing stable (solid lines) and unstable (dash lines) solutions show remarkable agreement with experiment. A flattening of the tunability curve close to 16.2 MHz appears in all three branches, showing interaction of the oscillator with a different mechanical mode in the beam. Inset: Detail of such flattening for the third branch. **d,** Inset: phase noise at 1kHz offset for our PFO as a function of $\phi$. Little dependence is observed except in the proximity of the flattening feature shown in **c**. **d,** Phase noise measurements for our PFO in both a standard case and at the optimum phase value, showing a reduction of the noise. For comparison, the phase noise of a standard direct-drive oscillator is shown for the same oscillator energy, indicating higher phase noise than PFO over most of the frequency range.

SYNOPSIS TOC

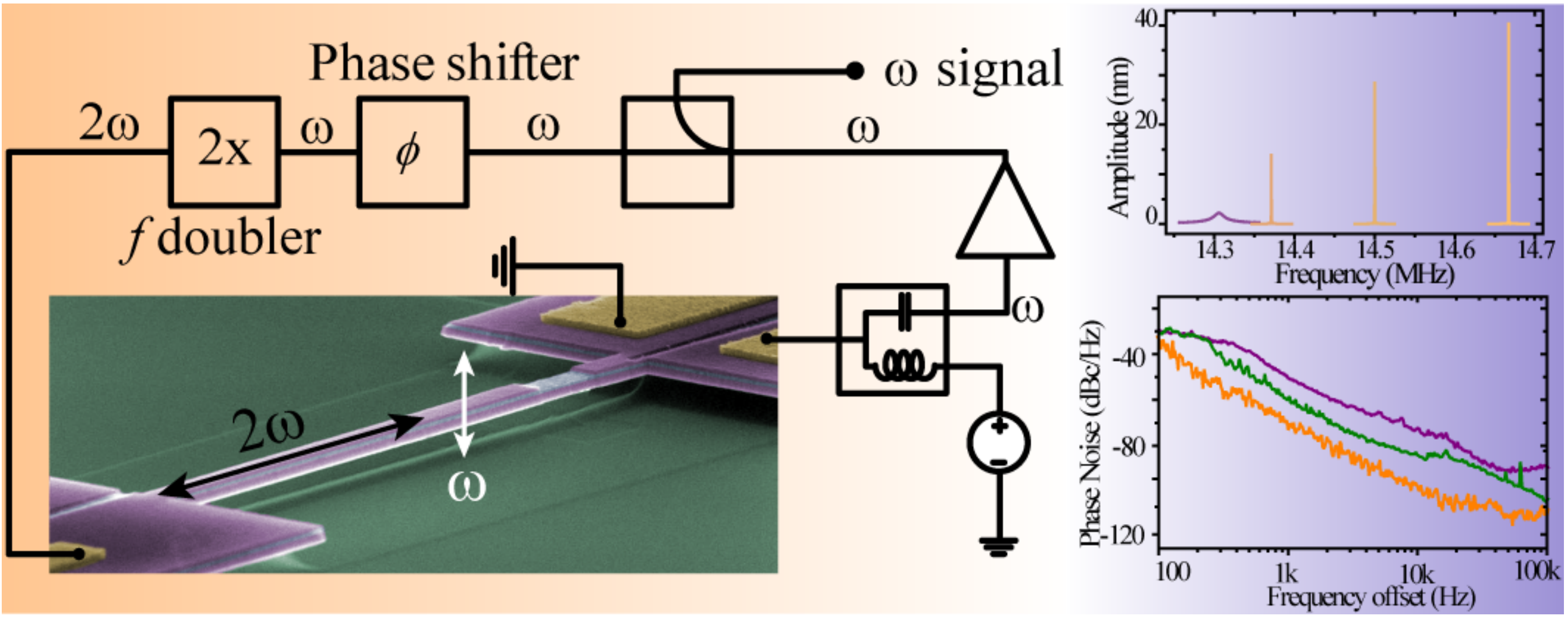